# Reversing the Pump-Dependence of a Laser at an Exceptional Point

M. Brandstetter[1,2], M. Liertzer[3], C. Deutsch[1,2], P. Klang[2,4],

J. Schöberl[5], H. E. Türeci[6], G. Strasser[2,4], K. Unterrainer[1,2], S. Rotter[3]*

Affiliations:

[1]Photonics Institute, Vienna University of Technology, A-1040 Vienna, Austria.

[2]Center for Micro- and Nanostructures, Vienna University of Technology, A-1040 Vienna, Austria.

[3]Institute for Theoretical Physics, Vienna University of Technology, A-1040 Vienna, Austria.

[4]Institute for Solid-State Electronics, Vienna University of Technology, A-1040 Vienna, Austria.

[5]Institute for Analysis and Scientific Computing, Vienna University of Technology, A-1040 Vienna, Austria.

[6]Department of Electrical Engineering, Princeton University, Princeton, New Jersey 08544, USA.

*Correspondence to: stefan.rotter@tuwien.ac.at

## Summary

When two resonant modes in a system with gain or loss coalesce in both their resonance position and their width, a so-called "Exceptional Point" occurs which acts as a source of non-trivial physics in a diverse range of systems[1–4]. Lasers provide a natural setting to study such "non-Hermitian degeneracies", since they feature resonant modes and a gain material as their basic constituents. Here we show that Exceptional Points can be conveniently induced in a photonic molecule laser by a suitable variation of the applied pump[5]. Using a pair of coupled micro-disk quantum cascade lasers[6], we demonstrate that in the vicinity of these Exceptional Points the laser shows a characteristic reversal of its pump-dependence, including a strongly decreasing intensity of the emitted laser light for increasing pump power. This result establishes photonic molecule lasers as promising tools for exploring many further fascinating aspects of Exceptional Points, like a strong line-width enhancement[7–9] and the coherent perfect absorption of light[10,11] in their vicinity as well as non-trivial mode-switching[12] and the accumulation of a geometric phase[13] when encircling an Exceptional Point parametrically.

# Main Text

From the standard physics text book we know that modes in a closed resonator are conveniently described as the orthogonal eigenstates of a Hermitian operator or matrix, whose real eigenvalues are the corresponding mode frequencies or energies. In the case of resonators which are open or which feature internal loss or gain, the corresponding matrices are non-Hermitian, featuring complex eigenvalues and non-orthogonal eigenstates. The quasi-bound resonances described by such eigenstates can give rise to a so-called "Exceptional Point" (EP) which is a "non-Hermitian degeneracy" at which both the real and imaginary parts of two eigenvalues are identical, such that both the position and the width of two resonances are the same. Quite differently from a conventional, Hermitian degeneracy, not only the eigenvalues, but also the eigenvectors coalesce at an EP, leading to a whole host of interesting phenomena[1–4] which have recently attracted enormous interest[9–18]. In particular, experiments on parity-time (PT) symmetric systems[19,20], for which the EPs occur on the real frequency axis, have been a driving force behind recent progress[21–26]. Already in the first experimental demonstration of this kind[21], it was shown that a system of two passive waveguides radically changes its transmission, when being steered parametrically through an EP. Whereas earlier work with passive metal cavities[27] suggests that such a complete reversal of system properties close to an EP is not necessarily restricted to the case of PT-symmetry, it remains largely unexplored how to observe such a reversal in active optical structures with gain.

In particular, as the presence of gain can push a device across the lasing threshold, the question arises in which way EPs enter the emission characteristics of a laser. Lasing signatures of EPs have, indeed, already been discussed as, e.g., in terms of fast self-pulsations in the laser dynamics near an EP[7], a diverging laser line width right at the EP[7,8], as well as in relation to characteristic features in a laser's behaviour below or at threshold[10,11,18]. An important milestone still missing in the attempt to connect laser physics with EPs is an experimental demonstration of a similar reversal of lasing properties as realised for the passive systems discussed above. A theoretical proposal along these lines has recently been put forward by several of the authors[5]. Its key idea is to use the pump applied to a system of coupled lasers to steer the device through the vicinity of an EP. Since the theoretical details of how an EP enters the non-linear lasing equations[28] were already laid out in full detail in Ref. 5, we reduce these equations here to a simple linear form, where each laser mode is replaced by a resonant level. Restricting ourselves to only two modes *A*,*B*, which we assume to be pumped separately from each other, we end up with a 2x2 matrix description, in which the diagonal elements describe the levels *A*,*B* with frequencies *a*,*b* and effective gain or loss *α*,*β*. The off-diagonal elements *γ* stand for their coupling,

$$H = \begin{pmatrix} a+i\alpha & \gamma \\ \gamma & b+i\beta \end{pmatrix}.$$

Consider now the situation of two levels with identical frequencies (*a*=*b*), one of which (level *A*) experiences strong net-gain ($\alpha \gg |\gamma|$) and the second one (level *B*) featuring strong net-loss ($\beta \ll -|\gamma|$). For this case the non-Hermitian matrix *H* has one eigenvalue in the upper half of the complex plane and one in the lower half, corresponding to one mode above and the other mode below the lasing threshold (as given here by the real axis). If, in a next step, we continuously add more gain to level *B* until both levels have the same gain (i.e., we sweep $\beta \to \alpha$), we find the following curious behaviour (see Fig. 1a): First the two eigenvalues approach each other such that the lasing level is pulled towards the lasing threshold (corresponding to a decrease of light emission). After passing the EP, where the two eigenvalues coalesce, this behaviour changes abruptly and the eigenvalues are repelled in their real parts and then both shifted beyond the lasing threshold (corresponding to an increase of light emission). We may thus conclude that such a device displays exactly the desired reversal of system properties associated with an EP. Note that this behaviour is robust in the sense that it also survives a slight detuning between the two level frequencies, for which case the eigenvalues just pass the vicinity of an EP (see Fig. 1b).

A natural experimental arena in which the above lasing effect could be realised is given by coupled ridge or microcavity lasers[6,29] in each of which one level can be brought to lase. A direct implementation of such a "photonic molecule laser", however, meets a number of challenges: First of all, the coupled laser needs to operate in the single-mode regime, as the onset of additional modes would give rise to nonlinear mode-competition effects which may overshadow the effect of an EP. Secondly, in the absence of pump, the resonator needs to be strongly absorbing such as to achieve the initial gain/loss-configuration from which the above pump sweep ($\beta \to \alpha$) starts off. Thirdly, the laser geometries as well as the coupling gap between them need to be engineered on the scale of the lasing wavelength such as to obtain sufficiently similar resonator frequencies as well as the right coupling strength.

Systems which we found to fulfil all of the above stringent criteria are photonic molecule quantum cascade lasers operating in the THz regime[6,30]. Their emission wavelength (~100 µm) is comparable to their size such that these devices feature both a stable regime of single mode operation as well as a much higher fault tolerance with respect to geometrical imperfections as compared to corresponding lasers emitting in the visible spectrum of light. Furthermore, quantum cascade lasers are electrically pumped devices, they provide a symmetric Lorentzian gain profile[31] and do not suffer from surface recombination. Specifically, we fabricated pairs of disk shaped lasers which we placed in close vicinity to each other in order to achieve sufficiently strong mode coupling. The active region of the laser is sandwiched, on top and at the bottom, by two metal layers which act both as a waveguide and as a contact for electrically pumping the device. Due to their finite conductivity, these metal layers provide much of the required loss already quite naturally. Since, however, even higher loss values are necessary to observe an EP, we reduced the thickness of the active region and added an additional absorption layer (see Supplementary sections S1 and S2). Figure 2a shows an image of a fabricated device.

The pump-dependence of this photonic molecule laser is shown in Fig. 2b, where the emitted laser light intensity (see false colour plot) is displayed as a function of the bias fields $F_A$, $F_B$ applied to disk A and B, respectively. In this plot the specific pump-trajectory which induces an EP along the discussion above, is realised both as a vertical and as a horizontal line (see insets). In both of these configurations the starting point is such that the applied bias is above threshold in one disk and below in the other disk. When gradually adjusting the lower field value to the higher value, the laser, indeed, shows the characteristic behaviour predicted above: After a regime where an increase of the bias field does not influence the output intensity at all, we first observe a strong reduction in the emitted light intensity, followed by an increase beyond the initial value. The observation of a reversal in the laser's pump dependence constitutes a clear signature of the presence of an EP near those parameter values where the reversal occurs. To rule out that the observed behaviour is caused by some other mechanism, we carried out a number of additional checks.

As a first test, we performed extensive numerical simulations of the studied setup to verify the presence of the EP in the complex eigenvalue surfaces explicitly. Since the coupling between two disks is spuriously overestimated in 2D scalar calculations, we performed 3D vectorial simulations of the photonic molecule device. To emulate the effect of the varying pump strength in the experiment, we solved the Helmholtz equation for the resonances of the photonic molecule under variation of the imaginary part of the index of refraction $n_A$, $n_B$ of the two respective micro disks. The first insight derived from these calculations is that the modes which start lasing first in the experiment are whispering gallery modes with radial quantization number n=3 (see inset in Fig. 3 and Supplementary section S3). For these modes, a variation of the imaginary part of the refractive index in one disk, e.g., $\text{Im}(n_B)$, yields exactly the expected avoided level crossing (see Fig. 1c) as previously obtained with the 2x2 matrix model (see Fig. 1b). When varying the imaginary parts of both refractive indices, $\text{Im}(n_A)$, $\text{Im}(n_B)$, we obtain the same characteristic dependence on these parameters (see Fig. 2c) as observed in the experiment when varying the applied field strengths $F_A$, $F_B$ (see Fig. 2b). As a final test, we also successfully verified in Fig. 3, that our experiment reproduces the expected real frequency shift induced by the EP. As already suggested by the plots in Fig. 1, we observe that the mode frequencies can be shifted to higher as well as to lower values when passing the EP (see Figs. 3a,b). The spectral data contained in Fig. 3 also clearly demonstrates that the entire recorded pump sweep across the EP involves just a single lasing line which is switched off and then on again.

In summary, we have unambiguously identified the influence of an EP on the operation characteristics of a photonic molecule laser by a suitable variation of the applied electrical pump. We observe very counter-intuitive lasing effects which result from the movement of complex eigenvalues in the vicinity of the EP. Our results provide a solid basis for many further studies on how the fascinating aspects of EPs can be "brought to light" with lasers.

## Methods Summary

The quantum cascade laser active region consists of a GaAs/Al$_{0.15}$Ga$_{0.85}$As heterostructure, grown with molecular beam epitaxy.[32] In order to improve the electrical contact, highly doped GaAs layers are grown on top and bottom of the active region, sandwiched by a double metal waveguide.[33] We reduced the thickness of the structure to increase the waveguide losses using a dry-chemical process by reactive ion etching. The devices are mounted to a copper heat sink with indium and contacted using a wire-bonding technique. All measurements are performed in continuous wave (cw) mode in order to achieve stable operation conditions. For the integral measurements of the bias dependent intensity plots in Fig. 2 the samples are mounted on a probe which is put directly in a liquid helium dewar to efficiently cool the devices and to obtain a reproducible and stable temperature. The emitted intensity is measured using a Ga doped Ge detector. The spectral measurements are performed using a Brucker Vertex 80 FTIR spectrometer with a resolution of 2.25 GHz. The emitted light is recorded with an attached pyroelectric deuterated triglycine sulfate (DTGS) detector.

To simulate the photonic molecule laser numerically we solved the vectorial Helmholtz equation in 3D, using the high order finite element package *ngsolve*[34] developed by one of the authors (J.S.). The dimensions of the device and the mode wave lengths were faithfully reproduced on the discretization grid. To account for the outgoing flux, we implemented a perfectly matched layer around the device. The modes lasing in the experiment were determined by a direct comparison between the measured and the calculated frequency spectra which were found to agree very well.

**Supplementary information** is linked to the online version of the paper.

**Acknowledgements** The authors would like to thank the following colleagues for very fruitful discussions: A. Benz, A. Cerjan, S. Esterhazy, G. Fasching, L. Ge, D. O. Krimer, M. Janits, K. G. Makris, M. Martl, J. M. Melenk, L. Nannen, A. Regensburger, and A. D. Stone. Financial support by the Vienna Science and Technology Fund (WWTF) through Project No. MA09-030, by the Austrian Science Fund (FWF) through Projects No. F25 (SFB IR-ON), No. F49 (SFB NextLite), by the NSF Grant No. EEC-0540832 (MIRTHE), and by the DARPA Grant No. N66001-11-1-4162 is gratefully acknowledged. We are also indebted to the administration of the Vienna Scientific Cluster (VSC) for granting us free access to computational resources.

**Author Contributions** M.B. and C.D. performed the experiments, M.L. and J.S. carried out the numerical calculations. P.K. and G.S. grew the quantum cascade hetero-structure. S.R., K.U. and H.T. initiated and supervised the project at all stages. M.B., M.L. and S.R. wrote the first manuscript draft. All authors discussed the results and commented on the manuscript.

**Author Information** The authors declare no competing financial interests. Correspondence and requests for materials should be addressed to S.R. (stefan.rotter@tuwien.ac.at).

# Figure Captions

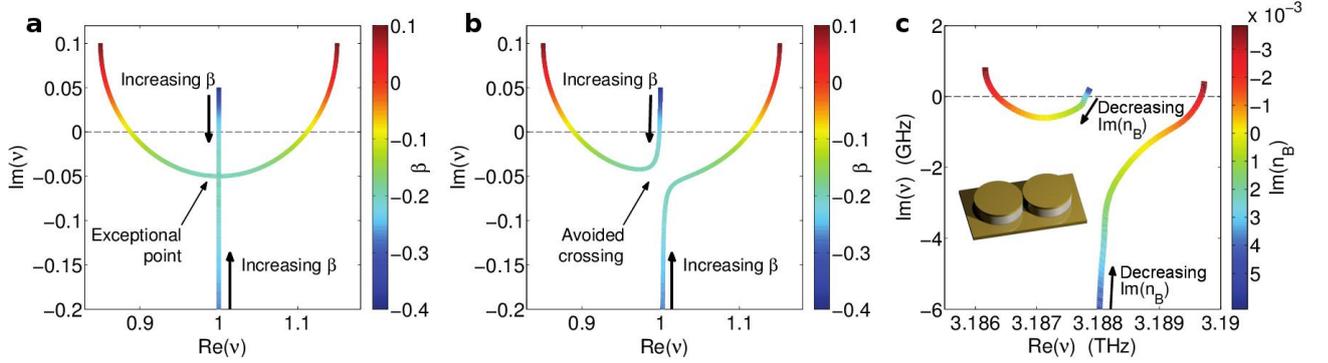

**Figure 1: Movement of complex eigenvalues in the vicinity of an EP.**

The behaviour of a laser with two coupled modes can qualitatively be understood by the eigenvalues of the 2x2 matrix $H$ given in the text. In **a,** the parametric dependence of these eigenvalues is shown for zero energy splitting between the two modes ($a=b=1$) and coupling strength $\gamma=0.15$. When one level is strongly amplified ($\alpha=0.1$) and the other level is swept from being strongly attenuated ($\beta = -0.4$) to being strongly amplified ($\beta \to \alpha$), the two eigenvalues first approach each other until they coalesce in an EP. After passing the EP, they are repelled in their real parts and shifted upwards in the complex plane. The real axis (dashed line) represents the lasing threshold and whenever it is crossed by an eigenvalue from above (below) a laser mode turns off (on). **b,** This characteristic reversal of the eigenvalue movement is also seen for a finite frequency splitting ($a=1.0$, $b=1.002$), leading to an avoided eigenvalue crossing in the complex plane. **c,** Numerical results for the complex resonance frequencies of the experimental device obtained with 3D finite-element calculations. The movement of resonance frequencies under variation of the imaginary part of the index of refraction in disk B shows the same avoided level crossing as in **b**. An almost degenerate mode pair with a similar eigenvalue movement has been omitted for clarity.

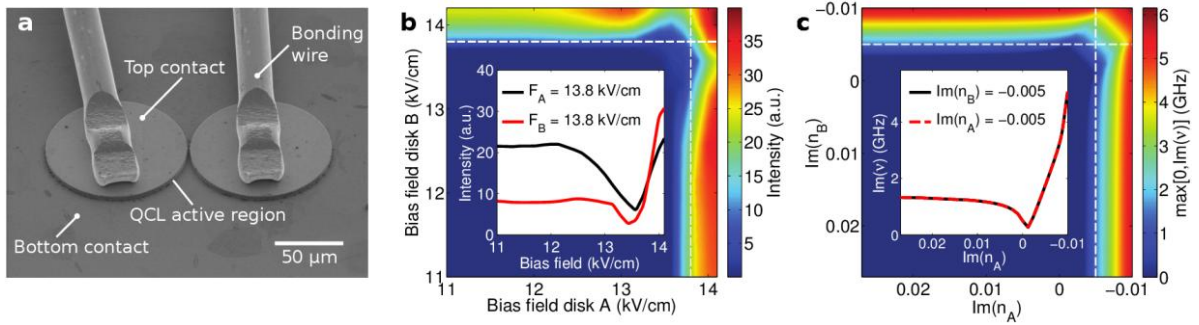

**Figure 2: Photonic molecule laser and its pump dependence.**

**a,** Image of the studied photonic molecule quantum cascade laser taken with a scanning electron microscope (disk radius $r=47$ μm, height $h=3.5$ μm, inter-cavity distance $d=2$ μm). **b,** Measured intensity output from a photonic molecule laser (integrated over all frequencies and emission directions) as a function of the electric field strength applied to the two individual disks (in the dark blue region the laser is off). The upper right corner contains the non-monotonic pump-dependence expected for an EP. When the field strength in one of the disks is fixed and the other disk is steered through the EP's vicinity (see white dashed lines) this results in a characteristic reversal of the laser's pump dependence (see inset for the corresponding intensity curves). **c,** Numerical results from the 3D simulations: The maximum of the positive imaginary parts of the complex resonance frequencies is shown as a function of the amplification in each disk, as given by the corresponding imaginary parts of the refractive indices (only the $n=3$ modes are considered which are lasing in the experiment). Note the excellent agreement which we find between these calculations and the experimental data in **b**.

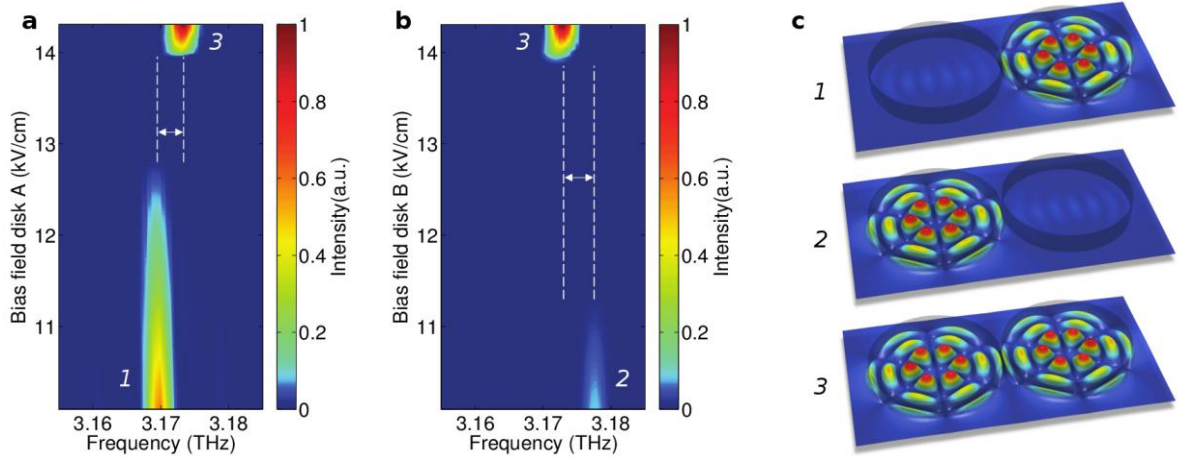

**Figure 3: Emission spectrum and lasing modes.**

**a,** Experimental emission spectrum recorded with the device in Fig. 2**a**, as a function of the pump strength in disk A with disk B at a constant pump value slightly above threshold. The colour gradient is proportional to the measured intensity and shows that along this pump sweep a single lasing line first turns off and then re-emerges with a shifted frequency (see frequency splitting indicated by the double-arrow). Note that the measured line width is limited here by the resolution of the spectrometer. **b,** When varying instead the pump strength in disk B with disk A slightly above threshold, the equivalent behaviour is observed, however, with a frequency shift of opposite sign. **c,** Electric field profiles ($|E(x)|$) of the involved modes as calculated at the bottom and top end of the pump sweeps shown in **a,b**.

# Methods

**Active region and fabrication**

The quantum cascade laser active region consists of a GaAs/Al$_{0.15}$Ga$_{0.85}$As heterostructure, grown with molecular beam epitaxy (MBE). It is based on a 3 well LO phonon depletion design[32] with a designed emission frequency of 3.2 THz. The layer sequence starting from the injector barrier in nanometers is **4.8**/9.6/**2**/7.4/**4.2**/16.1, the 16.1 nm well is homogeneously n-doped with a density of 7.5 x 10$^{15}$ cm$^{-3}$. The QCL module is repeated 271 times giving a total thickness of 13 µm. In order to improve the electrical contact, highly doped GaAs layers were grown on top and at the bottom of the active region with a thickness of 50 nm and 100 nm respectively. For the fabrication of the devices in the double-metal waveguide[33] we covered the active region and a carrier substrate with a 1 µm thick gold layer by magnetron sputtering. In the subsequent Au-Au thermo-compression bonding step we attach the active region to the carrier substrate. Then the GaAs substrate, on which the QCL was grown onto, is removed wet chemically.

We reduced the thickness of the structure to increase the waveguide losses using a dry-chemical process by reactive ion etching (RIE), providing a smooth surface and homogeneous etch rates across the sample. We define the gold top contact/waveguide layer and an optional Ti absorption layer by a photo-lithography/lift-off process, acting as a self-aligned etch mask for the subsequent reactive ion etching process, defining the resonator. The devices are then mounted to a copper heat sink with indium and contacted using a wire-bonding technique.

**Measurement setup**

All measurements are performed in continuous wave (cw) mode in order to achieve stable operation conditions. For the integral measurements of the bias dependent intensity plots in Fig. 2 the samples are mounted on a probe which is put directly in a liquid helium dewar to efficiently cool the devices and to obtain a reproducible and stable temperature, which is important for cw operation. The emitted intensity is measured using a Ga doped Ge detector. The spectral measurements are performed using a Brucker Vertex 80 FTIR spectrometer with a resolution of 2.25 GHz. The emitted light is recorded with an attached pyroelectric deuterated triglycine sulfate (DTGS) detector. The sample is mounted in a liquid helium cooled flow cryostat, attached to the spectrometer.

**Numerical calculations**

The numerical results are obtained by solving the fully vectorial three-dimensional Helmholtz equation,

$$\left(\nabla \times \nabla \times + n^2 k_m^2\right) \vec{E}_m(\vec{x}) = 0,$$

with open (outgoing) boundary conditions. The real part of the index of refraction *n=3.61* has been determined by comparing theoretical calculations and experimental measurements of devices featuring multiple laser modes. The imaginary part of the index of refraction is varied as shown in Figs. 1 and 2. The metal waveguide layers were approximated as perfect conductors and the outgoing boundary conditions are imposed using a perfectly matched layer. The dimensions of the simulated photonic molecule are the same as in the experiment. The calculations were carried out using the open source library *ngsolve*[34], which was developed by one of the authors (J.S.) based on a high order finite element discretization technique.

# Supplementary Information

## S1 Thinned THz QCL active regions

A thinner active region provides higher waveguide losses (see section S2) as required for the observation of an EP. Decreasing the device height also reduces the electrically generated heat, thereby suppressing thermal effects on the lasing behaviour in cw operation mode. In turn, decreasing the active region/waveguide thickness of THz QCLs with double metal waveguide leads to slightly increased threshold currents due to higher losses[1] but no influence due to microcavity effects are expected[2]. In Fig. S1 the experimental bias field/intensity vs. current density plots of devices with different active region thicknesses are shown. For the un-thinned device with 13 µm active region thickness, indicated by the black curves, the highly doped top contact layer is not removed. The blue curve shows the characteristics of a 3.5 µm thick device, where an additional Ti absorption layer is introduced between the bottom contact and the active region, similar to the structures used for the devices in the main text. The threshold field of the 13 µm device is lowered when thinning the device height since in the thinning process the top GaAs contact layers are removed, leading to additional contact voltages. Furthermore the threshold current density increases for thinner devices due to increased waveguide losses.

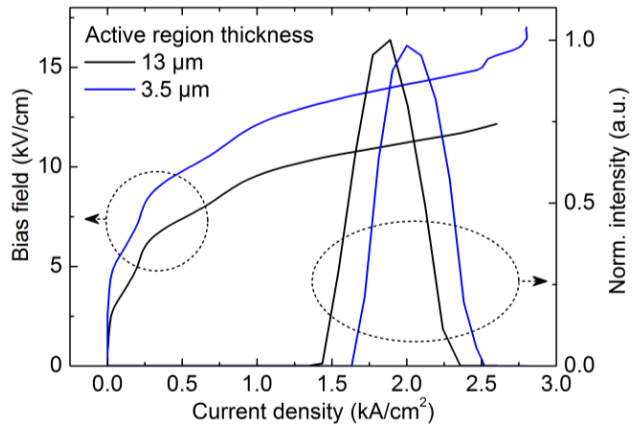

**Figure S1. Changing the losses by tuning the active region thickness.**

Measured data for the bias field/emitted intensity vs. current density plotted for single disks with different active region thickness. The black curve indicates a 13 µm thick device including an n-doped GaAs top contact layer. The blue curve shows the characteristics for a device which is thinned to 3.5 µm including an additional Ti absorption layer.

## S2 Loss calculation of double-metal waveguides

For the calculation of the waveguide loss in a double-metal structure we make use of a 1D waveguide solver based on a transfer-matrix formalism. Since the QCL structure consists of 25% $Al_{0.15}Ga_{0.85}As$, we model the active region using the mean value of the complex refractive indices, with a mean doping density of $5.3 \times 10^{15}$ $cm^{-3}$. The 50 nm thick GaAs contact layer on the bottom of the active region is assumed to have a doping density of $5 \times 10^{18}$ $cm^{-3}$. The simulation model consists of the active region stacked between two gold contact layers and a Ti absorption layer between the active region and the top contact layer. The material parameters of Ti are described using a Drude-model with a plasma frequency $\omega_p = 2\pi \cdot 6.09 \times 10^{14}$ s$^{-1}$ and a damping frequency of $\omega_t = 2\pi \cdot 1.14 \times 10^{13}$ s$^{-1}$, the parameters for gold are taken as $\omega_p = 2\pi \cdot 2.18 \times 10^{15}$ s$^{-1}$ and $\omega_t = 2\pi \cdot 6.44 \times 10^{12}$ s$^{-1}$ [3].

Figure S2A shows the calculated waveguide loss at varying Ti absorption layer thickness for waveguides with 3.5 µm thickness. The waveguide losses are increasing with the absorption layer thickness. Figure S2B depicts the waveguide losses vs. active region thickness at 90 nm and 5 nm thick absorption layers. The losses are higher for thinner devices due to a larger influence of the metal layers. The dashed lines indicate the parameters used in the experiment. A certain thickness of the active region is necessary since the coupling strength of the disks is reduced for thinner waveguides, which can be attributed to the small aperture and consequently to the strong diffraction of the emitted light. Thus an absorption layer is needed to achieve the right amount of losses and strong enough coupling simultaneously. Furthermore the active region thickness is chosen small enough to keep the electrical heat generation (due to the driving current) as small as possible, which is crucial for cw operation.

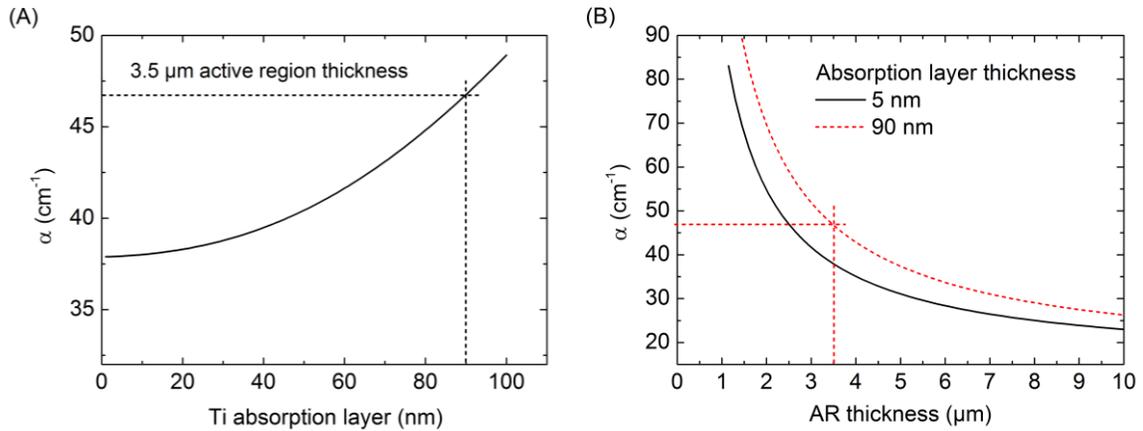

**Figure S2. Calculated waveguide losses.**

(A), The losses are increasing with increasing Ti absorption layer thickness. The dashed lines indicate the absorption layer thickness used in the experiment. (B), Waveguide losses for varying active region/waveguide thickness. For thinner cavities the influence of the metal layers is larger, providing higher losses. The used active region thickness is highlighted by the dashed line.

### S3  Mode identification

In order to identify the modes in the experimentally obtained spectrum we compare the measured spectra to a 3D simulation using a finite element solver. A refractive index of the active region of $n_r = 3.61$ is assumed and the waveguide layers are modeled as perfectly conducting. Due to the unequal spectral spacing of modes with different number of radial maxima we can identify modes by comparing the relative distances between their frequencies. Furthermore, coupled disk modes with fewer radial peaks $n$ show a larger splitting between the symmetric and antisymmetric mode and thus provide additional information on their identity.

Figure S3A shows the measured spectrum of a pair of coupled microdisk lasers with radius $r = 52.5$ µm, $h = 4.5$ µm waveguide thickness and an inter-cavity distance $d = 2$ µm. The device shows multimode emission due to processing imperfections. Comparing the measured modes to the calculation in Fig. S3B, indicated by the grey circles, yields very good agreement, allowing us to identify the measured mode of the device in the main text featuring an EP ($r = 47$ µm, $h = 3.5$ µm, $d = 2$ µm) to be of type ($n = 3$, $m = 3$) with 3 radial maxima (indicated by the red circle).

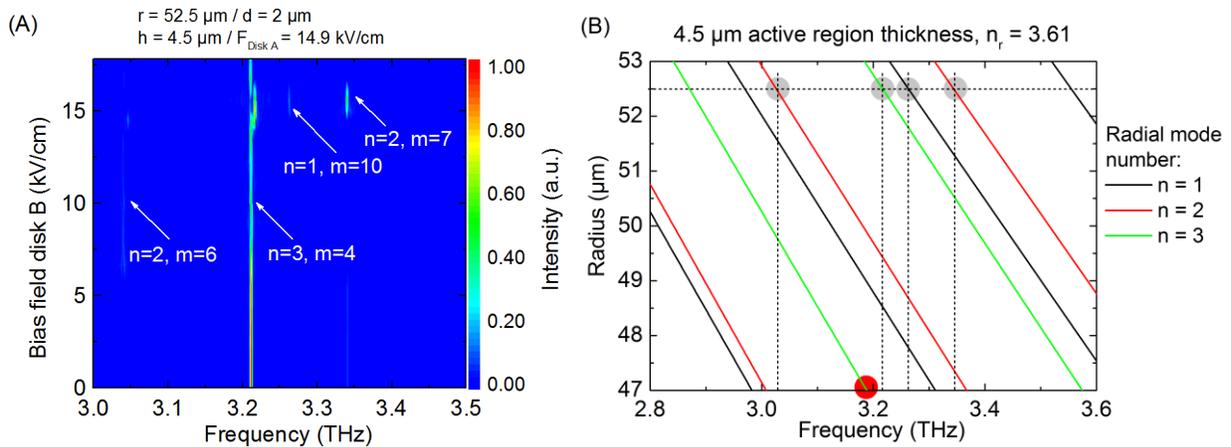

**Figure S3. Identification of the measured modes.**

(A), Spectra of a device showing multi-mode emission. The comparison to the simulation (B) allows us to identify the measured modes. We assume a refractive index of the active region of $n_r = 3.61$. The black, red and green lines indicate modes with $n = 1$, 2 and 3 radial maxima, respectively. The grey circles highlight the positions of the experimentally obtained modes, which perfectly fit to the simulation. The red circle indicates the mode occurring in the device presented in the main text.

We calculated the splitting of the coupled cavity modes with our 3D model, using the device dimensions from the experimental data. Figure S4 shows the splitting of mode arrangements with $n = 1$, 2, and 3 radial peaks. The dashed lines indicate the gap and the coupled cavity mode splitting for a device with dimensions used in the experiment ($r = 47$ µm, $h = 3.5$ µm). The calculated splitting of 3.3 GHz compares well to the measured value of 3.9 GHz. Since a lower radial mode number would lead to a significantly larger mode splitting, the calculation confirms our mode labeling. Using the spectral position and the size of the mode splitting we are thus able to identify unambiguously all the modes occurring in the experiment.

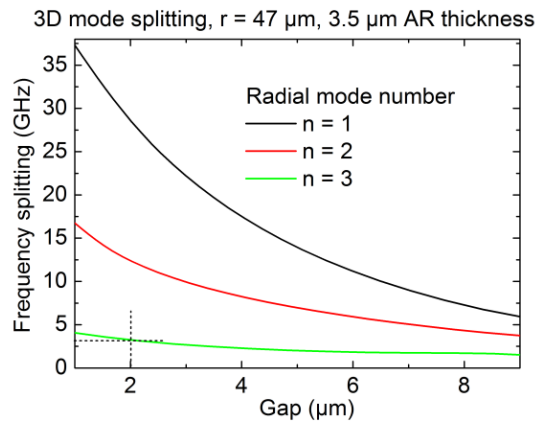

**Figure S4. Calculated splittings of coupled disk modes.**

Calculated splittings are shown for modes with $n = 1$, 2 and 3 radial maxima, following from the 3D simulation with dimensions as used in the experiment ($r = 47$ µm and $h = 3.5$ µm). The dashed lines indicate the splitting for a 2 µm gap. The experimental value of 3.9 GHz is in good agreement with the simulation.